%% file: xx.tex
\documentclass{JHEP3}\newcommand{\format} {\JHEPformat}

\usepackage{amsmath}
\usepackage{epsfig}
\usepackage{latexsym}

\newcommand{\JHEPformat} {
\bibliographystyle{JHEP}
\newcommand{\maketitlepage} {}
\abstract{\theabstract}
\keywords{\thekeywords}
\preprint{\thepreprint}
}

\newcommand{\TITLE}[1] {\newcommand{\thetitle} {#1}\title{#1}}
\newcommand{\ABSTRACT}[1] {\newcommand{\theabstract} {#1}}

\newcommand{\ADDRESS}[1] {\newcommand{\theaddress} {#1}}
\newcommand{\DATE}[1] {\newcommand{\thedate} {#1}\date{#1}}
\newcommand{\KEYWORDS}[1] {\newcommand{\thekeywords} {#1}}
\newcommand{\PREPRINT}[1] {\newcommand{\thepreprint} {#1}}

\def\half{\frac{1}{2}}

\def\pa{\partial}

\newcommand{\be}{\begin{equation}}
\newcommand{\ee}{\end{equation}}
\newcommand{\bea}{\begin{eqnarray}}
\newcommand{\eea}{\end{eqnarray}}

\def\a{\alpha}

\TITLE{Non-critical, near extremal  $AdS_6$ background 
as a holographic laboratory of four dimensional YM theory}

\ADDRESS{School of Physics and Astronomy\\
  The Raymond and Beverly Sackler Faculty of Exact Sciences\\
  Tel Aviv University, Ramat Aviv, 69978, Israel
}

\author{Stanislav Kuperstein, Jacob Sonnenschein\\
School of Physics and Astronomy\\
The Raymond and Beverly Sackler Faculty of Exact Sciences\\
Tel Aviv University, Ramat Aviv, 69978, Israel\\
E-mail:
\email{kupers@post.tau.ac.il, cobi@post.tau.ac.il}
}

\ABSTRACT{We study certain properties of the low energy regime of a
theory which resembles four dimensional 
YM theory in the framework of a  non-critical holographic gravity dual. We  use for the latter the 
near extremal $AdS_6$ non-critical SUGRA. We extract the glueball spectra that associates with the 
fluctuations of the dilaton, one form and the graviton and compare the results to those of 
the critical near extremal $D4$ model and lattice simulations. We show an area law behavior for 
the Wilson loop and screening for the 't Hooft loop. The Luscher term is found to be 
$-\frac{3}{24}\frac {\pi}{L}$.
We derive the Regge trajectories of glueballs associated with the 
spinning folded string configurations.   
}

\DATE{2004}

\KEYWORDS{Non-critical supergravity, AdS/CFT correspondence}

\PREPRINT{TAUP-2781-04\\{\tt hep-th/0411009}}

\format

\begin{document}

\maketitlepage


\section{ Introduction and Summary}

Ten dimensional type IIB superstring theory on $AdS_5\times S^5$ is holographically  equivalent to 
four dimensional ${\cal N}=4$  super Yang Mills theory. 
The $S^5$ transverse space with its $SO(6)$ 
isometry is required to dualize the R symmetry of the boundary gauge theory.
Using this logic the superstring dual of  ${\cal N}=1$ SYM should admit an $S^1$ as a transverse 
dimension and the dual of the  non-supersymmetric YM theory should be five dimensional with no transverse space.  
The fifth dimension plays the role of the re-normalization scale of the dual gauge theory.

Critical ten dimensional string (SUGRA) theories which are the
anti-holographic 
duals of confining gauge theories
with ${\cal N}=1$ supersymmetry are characterized by a KK sector of
states which are of 
the same mass scale as the 
hadronic states. There is no apparent way to disentangle 
these KK states from the hadronic states.  
This situation shows up for instance in the KS \cite{Klebanov:2000hb} 
and MN \cite{Maldacena:2000yy} models.

The KK modes combined with the argument about the R symmetry
serves as a motivation   to  study   non-critical string theories, 
as candidates for the string of QCD.
This idea was originally introduced in ~\cite{Polyakov:1998ju}
where a proposal of a dual of pure YM in terms of a 5d non-critical 
gravity background was made.  Following this paper  there were several attempts to 
find solutions of the non-critical effective action that are adequate as duals 
of gauge theories ~\cite{Klebanov:1998yy} ~\cite{Ferretti:1998xu}
~\cite{Ferretti:1999gj}
~\cite{Klebanov:1998yz} ~\cite{Garousi:1999fu} 
~\cite{Minahan:1999yr} ~\cite{Nekrasov:1999mn}
~\cite{Billo:1999nf} ~\cite{Armoni:1999fb} 
~\cite{Imamura:1999um} ~\cite{Ghoroku:1999bk}.

Recently, in \cite{Kuperstein:2004yk} we investigated 
the supergravity equations of motion associated with non-critical ($d>1$) type II 
string theories that incorporate RR forms.  Several  classes of solutions were derived.
In particular we found analytic backgrounds with a structure of 
$AdS_{p+2}\times S^{k}$ and numerical solutions that asymptote 
a linear dilaton with a topology of $R^{1,d-3}\times R \times S^1$. 
Unfortunately, for all these solutions the curvature in string units 
is proportional to $c=10-d$ and it cannot be reduced by taking a large $N$ limit 
like in the critical case. This means that the supergravity approximation
is not really valid. We conjectured, however, that the higher order corrections 
will modify the radii, while leaving the 
geometrical  structure of the background unchanged.
We also presented the AdS black hole backgrounds associated
with the $AdS_{p+2} \times S^k$ solutions.
In \cite{Polyakov:2000fk} the elevation of the SUGRA non-critical background into 
sigma models was discussed. Models with AdS target spaces
which have $\kappa$-symmetry and are completely integrable were derived.
Another important development in this program of deriving the string of QCD has been made 
in \cite{Klebanov:2004ya} where the 
$AdS_5\times S^1$ was derived from a non-critical SUGRA action 
that includes a term associated with a space filling
flavor brane.

The goal of this paper is to go one step forward and extract gauge dynamical  properties 
of confining theories from 
a non-critical SUGRA laboratory. In \cite{Kuperstein:2004yk} two classes of SUGRA 
backgrounds were shown to be duals of 
confining gauge theories: (i) near extremal AdS backgrounds (ii) backgrounds 
that  asymptote a linear  dilaton background. Since the latter were numerical solutions 
we choose to use the former option, namely 
a thermal AdS background.
To describe a four dimensional confining gauge theory we use the near extremal $AdS_6$ background.
The $AdS_6$ model, which is a member of the family of solutions mentioned above, 
includes a zero form field strength that may be associated with a 
$D4$ brane in a similar way to the $D8$ brane of the critical type IIA superstring theory.  
Near extremality, or the incorporation of a black hole, is achieved 
by   compactifying one of the $D4$ world-sheet coordinates on a circle and imposing 
anti-periodic boundary conditions. 
Some preliminary calculations have already been made in \cite{Kuperstein:2004yk}. 
In this paper we consider both SUGRA 
as well as semi-classical string computations. The first kind includes the extraction 
of the glueball spectra associated with the fluctuations of the dilaton, 
the graviton and the one form. Wilson loops, 't Hooft loops and spinning folded closed 
string are among the second kind.

An important question that is addressed in this paper is whether the results about the 
gauge dynamics that one derives from the non-critical models differ from those 
one extracts from the 
critical ones and in what ways. Here is  brief summary of the answer to this question.  

\begin{itemize}
\item
As advertised above, in comparison with the critical near extremal $D4$ brane model, the non-critical model
lacks the KK modes associated with the transverse $S^4$.
And, thus, in our case the corrections to everything we compute can be
large.
\item
The gauge theory dual of the non-critical SUGRA, is a theory in the large $N$ limit. However, unlike the 
usual limit of the AdS/CFT correspondence, here 't Hooft parameter is of order one  
$g^2 N\sim 1$ and not very large. 
\item
 Though the derivation of the glueball spectrum is similar to the one performed  in the critical 
 case there are certain differences in the resulting glueball spectra. For instance, 
 in the critical supergravity approach there is a degeneracy 
 between the $2^{++}$, $1^{++}$ and $0^{++}$ glueball states. This degeneracy is not present in our 
 results. On the other hand, we find that $M(2^{++})/M(0^{-+})=1$ is
 similar to  the critical 
 supergravity result $M(2^{++})/M(0^{-+}) \approx 1.2$ and the
 lattice result $M(2^{++})/M(0^{-+}) \approx 1.08$ .  

\item
The classical string configurations that correspond to  the  Wilson
loop and the 't Hooft loop admits the same behavior in the ten dimensional and six
dimensional string models, namely, area law behavior for the former and 
screening for the latter. However  the corresponding string tension behaves like 
$T_s\sim \sqrt{g^2 N} T^2$ and $\sim  T^2$ respectively, where $T$, 
the ``temperature'' is the inverse of the circumference of the
compactified $S^1$.  A similar difference occurs also for the 
tension of the folded spinning closed strings that admit a Regge trajectory behavior.

\item
The quantum fluctuations around the above classical configurations
differ between the critical and non-critical models simply due to the 
fact that the number of bosonic (and presumably also fermionic) directions is different.

For instance the Luscher term that measures the quantum deviation from
the linear potential was shown to be $-\frac{7}{24}\frac {\pi}{L}$ in
the critical case where $L$ is the separation distance between the corresponding quark and anti-quark.
Assuming that the fermionic fluctuations
are massive due to the coupling to the RR fluxes, as is the case in the critical model, 
the bosonic contributions take the form of $-\frac{3}{24}\frac
{\pi}{L}$, which is closer to the result found in lattice simulations.

\item
The KK modes of the critical string model take part as intermediate states 
in amplitudes of hadronic states. Intermediate states in  non-critical models 
of the type we consider here do not include these KK modes. 
For instance it was shown in \cite{Sonnenschein:1999re} that the 
dominant mode in the correlator of two Wilson loops is a KK mode in the critical model. 
In the non-critical case it is replaced by a glueball mode. 
\end{itemize}

The organization of the paper is as follows.
In Section \ref{Gs} we present the non-critical supergravity solution.
Section \ref{Tgs} is devoted to the calculation of the glueball spectrum
in the gauge theory in the context of non-critical supergravity.
We find an expression for the WKB approximation to the spectrum,
then compare it with numerical results for the lowest states.
We also compare our results to the lattice $YM_4$ results.
Although the lattice calculations and the supergravity calculations 
are valid in different regions (weak and strong coupling respectively)
we find that there is a qualitative agreement between the results.
In Section \ref{TWl} we give the stringy description of the $4d$ Wilson line. As
expected the background satisfies the condition for the area law behavior.
We also compute  
the classical energy
of the Wilson line. 
In Section \ref{TtHl} we discuss the 't Hooft line. 
Similarly to the critical case the calculations in the non-critical
background show that there is a screening of magnetic charge in the gauge theory.
In Section \ref{Css} we analyze the equation of motion of the string sigma model
in the given background. We find a remarkably simple semi-classical dispersion relation 
for a closed rotating string configuration. This relation describes a 
Regge trajectory in the dual gauge theory.


\section{General setting}
\label{Gs}

The non-critical AdS black hole solution presented in \cite{Kuperstein:2004yk}
has the following form:

\begin{equation}                \label{eq:bh}
l_s^{-2} ds^2 =
   \left( \frac{u}{R_{AdS}} \right)^2 
      \left[ f(u) d\theta^2 -dt^2 + \sum_{i=1}^{p-1} dx_i^2 \right]  
    + \left(\frac{R_{AdS}}{u} \right)^2
                   \frac{du^2}{ f(u) }  
    + R_{S^k}^2 d \Omega_k^2,
\end{equation}
where $f(u)=\left(1 - \left( \frac{u_\Lambda}{u} \right)^{p+1} \right)$
is the thermal factor.We denote here the location of the horizon by
$u_\Lambda$ to indicate the fact that it determines the scale 
$\Lambda_{QCD}$ in the dual gauge theory.
Other authors denote it as  $u_0$ or $u_T$ or $u_{KK}$.  
The radii appearing in the metric are:

\begin{equation}           \label{eq:AdSSRadii}
R_{AdS} = \left( \frac{(p+1)(p+2-k)}{c} \right)^{1/2}
\quad \textrm{and} \quad
R_{S^k} = \left( \frac{(p+2-k)(k-1)}{c} \right)^{1/2}.
\end{equation}
For $(p,k)=(3,5)$ and $(p,k)=(5,4)$ this metric was considered in
\cite{Hashimoto:1998if} \cite{Csaki:1998qr}
(for the radii of the critical case),
where the calculation of the glueball spectra in $YM_{3}$ and $YM_{4}$ 
was performed using critical $d=10$ supergravity.
In this paper we will analyze the $(p,k)=(4,0)$ case in the context of non-critical supergravity.
This will provide an alternative "non-critical" description of $YM_{4}$.

Apart from the metric (\ref{eq:bh}) the background includes the constant dilaton:

\begin{equation}
e^{2 \phi_{0} } = \frac{2 c}{p+2} \frac{1}{Q^2}
\qquad
\textrm{with}
\qquad
c=10 - d = 8 -p -k.
\end{equation}
Here $Q$ is the RR flux.
The coordinate $\theta$ has to be periodic in order to avoid a 
conical singularity at $u=u_\Lambda$. The circle parameterized by $\theta$
shrinks smoothly to zero if the period is chosen to be:

\begin{equation}
\beta =\frac{4 \pi R_{AdS}^2}{(p+1) u_\Lambda}.
\end{equation}


\section{The glueball spectra}
\label{Tgs}

According to the AdS/CFT correspondence \cite{Witten:1998zw} in order to determine  
the glue-ball mass spectrum   we 
must solve the linearized supergravity equations of motion in the
background (\ref{eq:bh}). 
For calculation of the glue-ball spectra in the framework of the critical supergravity
see \cite{Csaki:1998qr}, \cite{deMelloKoch:1998qs}, \cite{Hashimoto:1998if}, 
\cite{Constable:1999gb}, \cite{Brower:1999nj},  \cite{Minahan:1998tm}
\cite{Brower:2000rp}, \cite{Brower:2000rp}, \cite{Caceres:2000qe}. 
In $d=6$ the supergravity excitation we are interested in are those of the dilaton,  
the metric and the RR one-form appearing in the non-critical action. 
We should note that the RR 3-form in $d=6$ is not 
an independent field since its 4-form field strength is 
Hodge dual to the 2-form field strength of
the RR 1-form.   
There is also a possibility to add the NS-NS $2$-form field perturbation,
but we will not do it in this paper.

It appears that perturbing the equations of motion in the 
string frame leads to a complicated mixing between the dilaton and the
metric modes. We therefore will switch to
the Einstein frame,
where the action takes the following form:

\begin{equation}       \label{eq:TheActionEf}
S = \int d^{d} x \sqrt{g}
               \left( \mathcal{R} - \frac{4}{d-2}(\pa\phi)^2 + 
               \frac{c}{\alpha^\prime} e^{\frac{4}{d-2} \phi} \right)
   -  \half \sum_{l=2,6}  \int e^{\frac{d-2l}{d-2} \phi} F_{l} \wedge \star F_{l}. 
\end{equation}
In the following we will perform the calculation for general $p$, while
substituting the relevant value $p=4$ ($d=6$) only in the final results.
The background solution of (\ref{eq:TheActionEf}) involves the AdS
black hole metric (\ref{eq:bh}), the constant dilaton  $\phi^{(0)}$
and the RR $(p+2)$-form, which plays
the role of the cosmological constant as we pointed out above.
We write the perturbed fields as $g_{MN}= g^{(0)}_{MN} + h_{MN}$ and
$\phi= \phi^{(0)} + \bar{\phi}$
and arrive at the following set of linearized equations of motion
for the metric, the dilaton and the RR $1$-form:

\bea    
\label{eq:MetricLem} 
\nabla_M \nabla_{N} h^L_L + \nabla^2 h_{MN} - 2 \nabla^L \nabla_{(M}h_{LN)} 
        &=& \frac{2(p+1)}{R_{AdS}^2}h_{MN},  
   \\
\label{eq:DilatonLem} 
\nabla^2 \bar{\phi} &=& \frac{(p+1)(p+2)}{R_{AdS}^2} \bar{\phi},  
   \\
\label{eq:1formLem} 
\nabla^M F^{(2)}_{MN} &=& 0,
\eea
where in the Einstein frame $R^2_{AdS}=\frac{1}{c}(p+1)(p+2) e^{-\frac{4}{p}\phi^{(0)}}$.
We see that the metric satisfies the linearized version of the usual 
Einstein equation in $p+2$ dimensions with a negative cosmological constant.
On the other hand the dilaton equation (\ref{eq:DilatonLem}) differs from the Laplacian
equation $\nabla^2 \bar{\phi}=0$ one obtains in the critical $AdS_5 \times S^5$ case.
The term on the r.h.s. of (\ref{eq:DilatonLem}) appears due to the non-critical term
in the action (\ref{eq:TheActionEf}). Remarkably this equation
is similar to the dilaton equation in the critical dimension, which
includes the contribution of the
non-zero KK modes on the compact transversal space.

The WKB approximation of the graviton spectrum in the AdS black hole background 
was thoroughly analyzed by \cite{Constable:1999gb} for
arbitrary $p$. 
Generally one may consider various polarization of the graviton.
This results in states that correspond to the $2^{++}$, $1^{++}$ and $0^{++}$ glueball spectra 
in the dual  gauge theory.
In this section we will quote the results of the WKB calculation
in the graviton case and will compare it to numerical results. 
We will also analyze the dilaton and RR $1$-form equations.
Exactly as in the 10d critical case we can assign the 
correct parity and charge conjugation eigenvalues to the 
corresponding glueball states by considering the coupling between the 
supergravity fields and the boundary gauge theory.
This coupling can be determined from a DBI action plus a WZ term
for a single D$4$-brane in the given background.
For instance, from the WZ part one finds the RR field $A_{(1)}$ 
with a leg along the compact $\theta$-direction
couples to the gauge field $F_{\mu\nu}$ through the term 
$\epsilon^{ijkl} A_\theta F_{ij} F_{kl}$. It means that this term
$(P,C) = (-,+)$ and hence $A_\theta \to 0^{-+}$.
Similarly, for the RR field polarized along the non-compact
world volume directions we obtain $A_\mu \to 1^{++}$.
We also have $\phi \to 0^{++}$, since the dilaton couples to 
to the operator $\phi F^2$.

\subsection{The dilaton and  the RR $1$-form spectra}

We start by introducing the ansatz for the the dilaton field
$\bar{\phi} = b(u) e^{ikx}$, where $b(u)$ depends only on the radial
coordinate and $k_\mu$ is a $p$-vector along the Minkowski part of the metric.
We then have $M^2=-k^2$ as the Lorentz invariant mass-squared  of the
dual operator. 
We will not discuss here the KK modes related to solutions with a
non-trivial dependence on the compact coordinate $\theta$.
Plugging this ansatz into the dilaton equation (\ref{eq:DilatonLem}) 
and using the explicit form of the metric (\ref{eq:bh}) for $k=0$ we obtain
a 2nd order differential equation for $b(u)$:

\begin{equation}    \label{eq:LindDil1}
\pa_u \left( \left( u^{p+2} -u \right) \pa_u b(u) \right) + 
   \left( M^2 R_{AdS}^4 u^{p-2} - (p+1)(p+2) u^{p}  \right)  b(u) =0 
.
\end{equation}
Here the last term appears due to the non-Laplacian part in the
equations of motion.
In order to put the equation into 
a Schr\"oedinger like form we follow \cite{Constable:1999gb}  and 
 re-define the wave function according to $b(u) =  \gamma(u) \xi(u)$
with:
 
\begin{equation}
\gamma(u) \equiv \sqrt{\frac{u-u_\Lambda}{u \left(u^{p+1}-u^{p+1}_\Lambda \right)}}
\end{equation}
and introduce a new radial coordinate defined by $u=u_\Lambda \left(1+e^y \right)$.
Now the equation (\ref{eq:LindDil1}) reduces to:

\begin{equation}   \label{eq:Schroedinger1}
- \xi^{\prime \prime}(y) + V(y) \xi(y) =0.
\end{equation}
where the effective potential takes the following form:

\FIGURE[t]{
 \label{Vdilaton20}
\centerline{\input{Vdilaton20.pstex_t}}
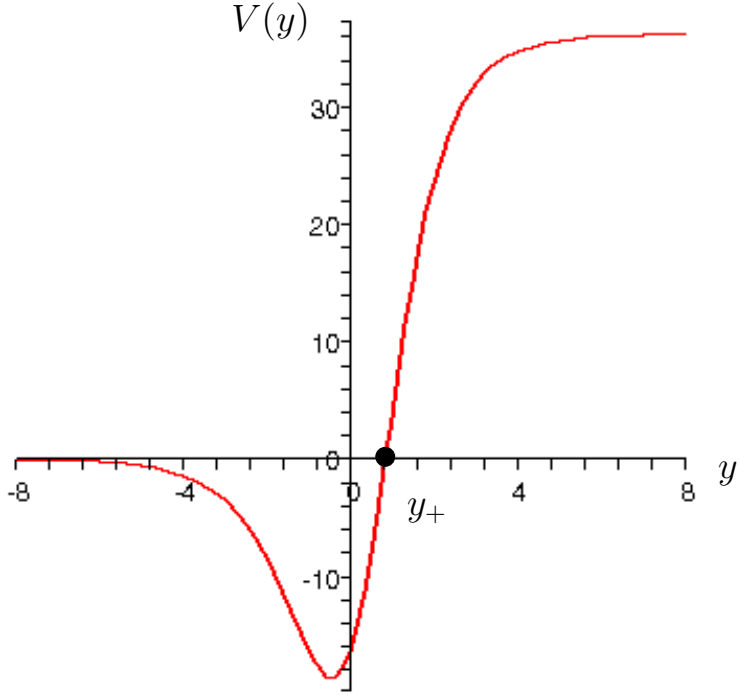
\caption{The effective potential (\ref{eq:DilPot}) for $p=4$ and
  $\frac{MR_{AdS}^2}{u_\Lambda}=20$.
  The plot demonstrates that there are two classical turning points at
  $y=y_+$ and at  $y=-\infty$.  
}}

\begin{eqnarray} \label{eq:DilPot}
V(y) &=&\frac{1}{4} + \frac{e^{2y} \left(p(p+2) (1+e^y)^{2(p+1)}-2p(p+2)(1+e^y)^{p+1} -1 \right)}
                        {4 (1+e^y )^2 ((1+e^y)^{p+1}-1)^2}   \nonumber \\
&&   - \left( \left( \frac{M R_{AdS}^2}{u_\Lambda} \right)^2 
                  - (p+1)(p+2)(1+e^y)^2 \right) \frac{e^{2y}(1+e^y)^{p-3}}{(1+e^y)^{p+1}-1}. 
\end{eqnarray}
Although the analytic expression (\ref{eq:DilPot}) is very
complicated, the potential still has a relatively simple shape
as it shown on Fig. (\ref{Vdilaton20}).

Before proceeding,  note that for $u \gg u_\Lambda$ 
(at large $y$) one has
\mbox{$\gamma(u) \approx \left( \frac{u}{R} \right)^{-(p+1)/2}$} and therefore 
for any function $\xi(y)$, which goes to zero at $y \to \infty$, 
the associated wave function $b(u)$ is normalizable
with respect to the $AdS_{p+2}$ metric measure.

Given the explicit form of the potential the mass parameter 
may be fixed be requiring that the Schr\"oedinger equation (\ref{eq:Schroedinger1})
produces a bound state at zero energy. Despite the complicated form of
the potential, it is, however, clear that there are no tachyonic (or $M^2=0$) modes in this case,
since for $M^2<0$ the potential $V(y)$ is positive everywhere.
Hence the corresponding operator in the gauge theory has a mass gap.

We will analyze the spectrum of the excitations applying the WKB
approximation following the method developed in \cite{Minahan:1998tm}.
This approximation is valid for large $M^2$, where the potential well is sufficiently deep.
The asymptotic
behavior of the potential (\ref{eq:DilPot}) is given by:

\bea
V(y \to \infty) &\approx&  \frac{1}{4}(p+1)(5p+9)
        - \frac{1}{2} ((p+1)(5p+9)-1) e^{-y} 
\nonumber \\
  && \qquad        + \left( \frac{3}{4} ((p+1)(5p+9)-1) 
                        - \frac{M^2 R_{AdS}^4}{u_\Lambda^2} \right) e^{-2y}+\ldots
\nonumber \\
V(y \to -\infty) &\approx&  \left(\frac{1}{4}(p+1)(5p+9)
          -  \frac{M^2 R_{AdS}^4}{(p+1) u_\Lambda^2} \right) e^y + \ldots.
\eea
It means that to the leading order in  $\frac{M R_{AdS}^2}{u_\Lambda}$ 
the classical turning points are:

\begin{equation} 
y_+ = \ln \left( \left( \frac{4}{(p+1)(5p+9)}\right)^{1/2} \frac{M R_{AdS}^2}{u_\Lambda} \right)
\qquad
\textrm{and}
\qquad
y_{-} = - \infty.
\end{equation}
Note that the point $y_-$ corresponds in terms of the original radial
coordinate to $u=u_\Lambda$.
In the WKB approximation the potential satisfies:

\begin{equation}  \label{eq:WKB}
\left(k-\half \right) \pi = \int_{y_-}^{y^+} \sqrt{-V(y)} dy,
\end{equation}
where $k$ is a positive integer. 
Using the method of \cite{Minahan:1998tm} we expand the integral 
as a series in powers of $\frac{u_\Lambda}{MR_{AdS}^2}$ leaving only the
terms appearing at $O(M)$ and  $O(M^0)$. 
To leading order in $M$ the integral (\ref{eq:WKB})
is approximated by:

\begin{equation}  
\int_{-\infty}^{y^+} \sqrt{-V(y)} dy
\approx  \frac{M R_{AdS}^2}{u_\Lambda}  \int_{-\infty}^{+\infty}  
                     \frac{e^{y}(1+e^y)^\frac{p-3}{2}}{\left((1+e^y)^{p+1}-1\right)^{1/2}} dy 
 \approx \frac{M R_{AdS}^2}{u_\Lambda} \frac{\sqrt{\pi} \Gamma \left(\frac{1}{p+1} \right)}
                            {(p+1)\Gamma \left(\half+\frac{1}{p+1} \right)} .
\end{equation}
There are two contributions to the next order term in the $1/M$
expansion.
The first contribution comes from integrating to $\infty$ instead of
$y_+$ in the leading order term. Therefore we should subtract from the
result the following contribution:

\begin{equation}  
 \frac{M R_{AdS}^2}{u_\Lambda}  \int^{\infty}_{y^+} 
                     \frac{e^{y}(1+e^y)^\frac{p-3}{2}}{\left((1+e^y)^{p+1}-1\right)^{1/2}} dy
\approx \half \left((p+1)(5p+9)\right)^{1/2}.
\end{equation}
The second contribution comes from the integration 
near $y=y_+$:

\begin{equation}  
 \int^{y^+} \left(
\sqrt{-V(y)}
 - \frac{M R_{AdS}^2}{u_\Lambda}  
                     \frac{e^{y}(1+e^y)^\frac{p-3}{2}}{\left((1+e^y)^{p+1}-1\right)^{1/2}} 
 \right)dy
\approx \half \left((p+1)(5p+9)\right)^{1/2} \left(1-\frac{\pi}{2} \right).
\end{equation}
Finally, adding up  and
solving for $M$ in terms of $k$ and substituting $p=4$
we obtain the masses of the spin-$0$ dilaton excitations:

\begin{equation}   \label{eq:WKBphi}
M^2_{\mathbf{0^{++}}, \phi} \approx  \frac{39.66}{\beta^2} k (k +5.02) + O(k^0)
\quad
\textrm{with}
\quad 
\beta = \frac{4 \pi R_{AdS}^2}{5 u_\Lambda}.
\end{equation}
We can also find the spectrum using the "shooting technique". Solving  
(\ref{eq:LindDil1}) numerically and matching the boundary condition at $u=u_\Lambda$
and $u=\infty$ results in a discrete set of eigenvalues of $M^2_k$.
Table \ref{TablePhi} compares the WKB expression (\ref{eq:WKBphi}) and the numerical 
results. We see a nice  agreement which improves for higher $k$.

\TABLE[tb]{
\label{TablePhi}
\begin{tabular}[b]{|c|r|r|}   
\hline   
 $k$  & WKB  & Numerical \\ 
\hline   
\ \ 1 & $15.45$  & $19.09$  \\   
\ \ 2 & $23.60$  & $26.14$  \\   
\ \ 3 & $30.89$  & $32.88$  \\   
\ \ 4 & $37.83$  & $39.47$  \\     
\ \ 5 & $44.58$  & $45.98$  \\
\ \ 6 & $51.21$  & $52.44$  \\ 
\hline
\end{tabular}   
\caption{Comparison of the $0^{++}$-glueball masses $M_{0^{++},\phi}$ in units of $\beta^{-1}$.
The WKB approximation is very close to the numerical results.}  
}

Next let us analyze the scalar glueballs related to the RR $1$-form $A_M$
directed completely along the compact $\theta$-coordinate.
We will consider the ansatz
$A_\theta = a(u) e^{ikx}$  with all other components vanishing
identically.
It will be useful to re-write the $1$-form equation of motion (\ref{eq:1formLem})
as:

\begin{equation}   \label{eq:1formEM}
\pa_M \left( \sqrt{g} g^{NK} g^{ML} (\pa_K A_L - \pa_L A_K) \right)=0.
\end{equation}
For our ansatz the only non-trivial equation occurs for  $N=\theta$.
The 2-nd order differential equation for
$a(u)$ reads:

\begin{equation}        \label{eq:a(u)}
\pa^2_u a(u) + \frac{p}{u} \pa_u a(u) + M^2 R_{AdS}^4
     \frac{u^{p-3}}{u^{p+1} - u_\Lambda^{p+1}} a(u ) =0.
\end{equation}
Following the same steps as in the dilaton case we end up with
the following effective potential:

\FIGURE[tb]{
 \label{Vrr5}
\centerline{\input{Vrr5.pstex_t}}
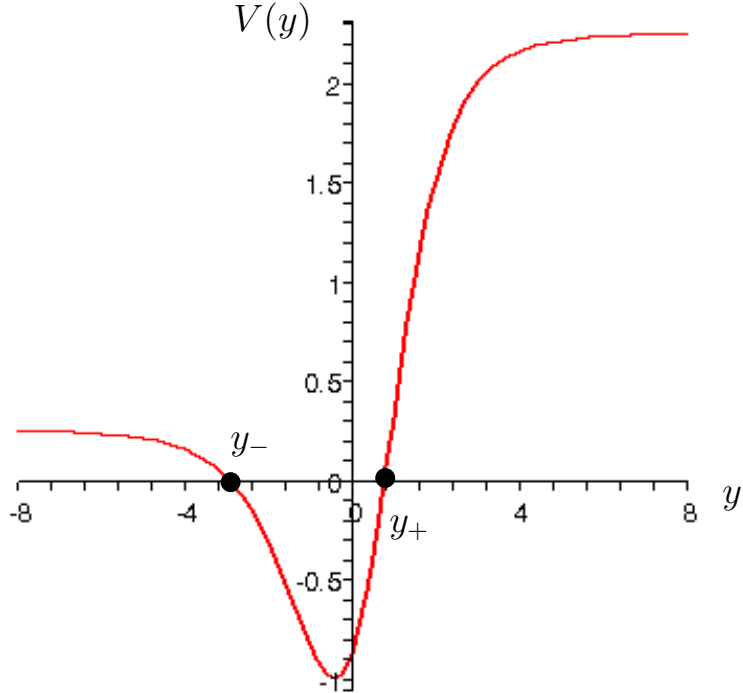
\caption{The effective potential (\ref{eq:RRPot}) for $p=4$ and
  $\frac{MR_{AdS}^2}{u_\Lambda}=5$.
  There are two classical turning points at
  $y=y_+$ and at  $y=y_-$.  
}}

\begin{equation} \label{eq:RRPot}
V(y)  = \frac{1}{4} + \frac{p(p-2) e^{2y}}{4(1+e^y )^2} 
   -  \left(\frac{M R_{AdS}^2}{u_\Lambda} \right)^2 
                     \frac{e^{2y}(1+e^y)^{p-3}}{(1+e^y)^{p+1}-1}. 
\end{equation}
The typical form of the potential is presented on Fig. (\ref{Vrr5}).
In this case the classical turning points are situated at:

\begin{equation}
y_+  \approx \ln \left( \frac{2 M R_{AdS}^2}{(p-3) u_\Lambda} \right)
\quad
\textrm{and}
\quad
y_-  \approx - 2 \ln \left( \frac{2 M R_{AdS}^2}{\sqrt{p+1} u_\Lambda}  \right).
\end{equation}
Note that in this case the inner
turning point is located away from the surface $u=u_\Lambda$.
Finally, the  spectrum of the glueballs related to the RR $1$-form
perturbation directed along the compact coordinate is:

\begin{equation}     \label{eq:WKBAtheta}
M^2_{\mathbf{0^{-+}}, A_{\theta}} \approx  \frac{39.66}{\beta^2} k \left(k +
  \frac{3}{2} \right) + O(k^0).
\end{equation}
We compare this result to the numerical solutions of (\ref{eq:a(u)}) in Table
\ref{TableA}. Again the agreement is very close.

\TABLE[tb]{
\label{TableA}
\begin{tabular}[h]{|c|r|r|}   
\hline   
 $k$  & WKB  & Numerical \\ 
\hline   
\ \ 1 & $9.96$   & $10.21$  \\   
\ \ 2 & $16.67$  & $16.81$  \\   
\ \ 3 & $23.14$  & $23.25$  \\   
\ \ 4 & $29.54$  & $29.62$  \\     
\hline    
\end{tabular}  
\qquad
\begin{tabular}[h]{|c|r|r|}   
\hline   
 $k$  & WKB  & Numerical \\ 
\hline   
\ \ 1 & $7.72$   & $7.44$  \\   
\ \ 2 & $14.08$  & $13.96$  \\   
\ \ 3 & $20.41$  & $20.32$  \\   
\ \ 4 & $26.72$  & $26.65$  \\     
\hline
\end{tabular}  
\caption{Comparison of the $0^{-+}$ glueball masses $M_{0^{-+}, A_\theta}$ (left)
and the $1^{++}$ glueball masses $M_{1^{++}, A_\mu}$ (right)
in units of $\beta^{-1}$.
For $0^{-+}$ masses the WKB approximation is close to the numerical results for any $k$.}  
}

Next let us consider the RR $1$-form with legs along the non-compact coordinates
$x_\mu$'s. Now the ansatz is $A_\mu = v_\mu \alpha(u) e^{ikx}$ 
with $v \cdot k =0$ and plugging this into (\ref{eq:1formEM})
we get:

\begin{equation}       \label{eq:b(u)}
\pa^2_u \alpha(u) + \frac{p u^{p+1} + u_\Lambda^{p+1}}{u \left( u^{p+1} - u_\Lambda^{p+1} \right)} 
                 \pa_u \alpha(u) + M^2 R_{AdS}^4 \frac{u^{p-3}}{u^{p+1} - u_\Lambda^{p+1}} \alpha(u ) =0.
\end{equation}
In this case the effective potential is:

\begin{eqnarray}
V(y)  &=& \frac{1}{4} + \frac{e^{2y} \left(p(p-2) (1+e^y)^{2(p+1)}-2(p^2+2)(1+e^y)^{p+1} -1 \right)}
                        {4 (1+e^y )^2 ((1+e^y)^{p+1}-1)^2} 
                        \nonumber \\
   &&  \qquad \qquad  -  \left(\frac{M R_{AdS}^2}{u_\Lambda} \right)^2 
                     \frac{e^{2y}(1+e^y)^{p-3}}{(1+e^y)^{p+1}-1} 
\end{eqnarray}
and the classical turning points are:

\begin{equation}
y_+  \approx \ln \left( \frac{2 M R_{AdS}^2}{(p-1) u_\Lambda} \right)
\quad
\textrm{and}
\quad
y_-  \approx - \infty.
\end{equation}
The WKB mass formula for the spectrum of the glueballs associated with 
RR $1$-form directed along the non-compact coordinates is:

\begin{equation}    
M^2_{\mathbf{1^{++}}, A_{\mu}} \approx  \frac{39.66}{\beta^2} k \left(k +
  \frac{1}{2} \right) + O(k^0).
\end{equation}
We compare this result to the numerical solutions of (\ref{eq:b(u)}) in Table
\ref{TableA}. Again the agreement is very close.

\subsection{The graviton spectra}

\TABLE[tb]{
\label{TableV}
\begin{tabular}[h]{|c|r|r|}   
\hline   
 $k$  & WKB  & Numerical \\ 
\hline   
\ \ 1 & $11.35$  & $12.57$  \\   
\ \ 2 & $18.36$  & $19.43$  \\   
\ \ 3 & $24.99$  & $25.99$  \\   
\ \ 4 & $31.49$  & $32.44$  \\     
\hline
\end{tabular}   
\caption{Comparison of the $1^{-+}$ glueball masses $M_{1^{-+}, h_{\theta\mu}}$ 
in units of $\beta^{-1}$.}  
}

The wave equations (\ref{eq:MetricLem}) for the metric fluctuations
about the $AdS_{p+1}$ 
black hole background
have been analyzed by \cite{Constable:1999gb}. Here we will use these results for $p=4$.
For different polarizations the linearized equation (\ref{eq:MetricLem}) 
reproduces a set of differential equations for spin-$2$, spin-$1$ and spin-$0$ glueballs.
The spin-$2$ part corresponds to the graviton polarized in the direction parallel
to the hyper-surface spanned by the world-volume coordinates $x_\mu$'s.
The appropriate ansatz for the metric in this case is:

\begin{equation}  \label{eq:polarization}
h_{ab} = \epsilon_{ab} \left(\frac{u}{R_{AdS}} \right)^2 H(u),  
\end{equation}
where $\epsilon_{ab}$ is a constant traceless polarization tensor  and 
the differential equation for the function $H(u)$ is:

\begin{equation}   \label{eq:Tde}  
\pa_u \left( \left( u^{p+2} - u^{p+1}_\Lambda u \right) \pa_u  H(u)\right) 
         - M^2 R_{AdS}^4 u^{p-2}  H(u) =0.
\end{equation}
The effective potential in derived from this equation is:

\begin{eqnarray} \label{eq:TPot}
V(y) &=&\frac{1}{4} + \frac{e^{2y} \left(p(p+2) (1+e^y)^{2(p+1)}-2p(p+2)(1+e^y)^{p+1} -1 \right)}
                        {4 (1+e^y )^2 ((1+e^y)^{p+1}-1)^2}   \nonumber \\
&&   -    \left( \frac{M R_{AdS}^2}{u_\Lambda} \right)^2   
                    \frac{e^{2y}(1+e^y)^{p-3}}{(1+e^y)^{p+1}-1}. 
\end{eqnarray}
It was pointed out in \cite{Constable:1999gb} that 
this effective potential is identical to the potential
one obtains for a minimally coupled scalar ($\nabla^2 \phi =0$). Based on this observation
the authors concluded that there is a degeneracy between the tensor and the scalar excitation.
As we see, however, in the framework of non-critical 
supergravity the dilaton is not minimally coupled,
but rather satisfies the equation (\ref{eq:DilatonLem}), and therefore the effective potentials 
(\ref{eq:DilPot}) and (\ref{eq:TPot})
are different.

Following the same steps as in the scalar glueball case we can find the 
WKB expression for the spectrum and compare it to the numerical results obtained 
directly from (\ref{eq:Tde}).
The final result is:

\begin{equation}   \label{eq:WKBT}
M^2_{\mathbf{2^{++}}, h_{\mu \nu}} \approx  \frac{39.66}{\beta^2} k \left(k +
  \frac{3}{2} \right) + O(k^0).
\end{equation}
Surprisingly this expression is identical to the result  (\ref{eq:WKBAtheta}) 
for the $0^{-+}$ glueballs in
the previous subsection. Moreover, 
this degeneracy holds also beyond the WKB approximation, since
the numerical calculations also produce the same spectrum.
This is quite unexpected, since
there is no redefinition of the function $H(u)$ or/and of the radial coordinate $u$,
that brings the differential equation (\ref{eq:Tde}) to the form (\ref{eq:a(u)})! 
We will return to this result in the end of the section, while comparing 
our results to the lattice calculations.

\TABLE[b]{
\label{TableS}
\begin{tabular}[h]{|c|r|r|}   
\hline   
 $k$  & WKB  & Numerical \\ 
\hline   
\ \ 1 & $9.96$   & $6.34$  \\   
\ \ 2 & $16.67$  & $15.58$  \\   
\ \ 3 & $23.14$  & $22.43$  \\   
\ \ 4 & $29.54$  & $29.01$  \\     
\hline
\end{tabular}   
\caption{Comparison of the $0^{++}$ glueball masses $M_{0^{++},h_{\theta\theta}}$ 
in units of $\beta^{-1}$.
The WKB approximation is close to the numerical results only for $k>1$.}  
}

Finally, let us quote the results for the spin-$1$ and spin-$0$ modes.
The solution that appears as vector in the gauge theory
is given by the ansatz (\ref{eq:a(u)}) with the polarization tensor 
satisfying:

\begin{equation}
\epsilon_{\theta \mu} = v_{\mu},
\qquad
\textrm{where}
\qquad
k \, \cdot \, v =0 
\quad
\textrm{and}
\quad
v^2=1.
\end{equation}
The ansatz is consistent with the equation of motion 
provided that $H(u)$ satisfies:

\begin{equation}   \label{eq:Vde}
\pa^2_u H(u) + \frac{(p+2)}{u} \pa_u H(u) + 
   \frac{M^2 R_{AdS}^4 u^{p-3}}{ \left(u^{p+1}-u_\Lambda^{p+1} \right)} H(u) =0.
\end{equation}
The effective potential in this case is given by:

\begin{equation} 
V(y)  = \frac{1}{4} + \frac{p(p+2)e^{2y}}{4(1+e^y )^2} 
   -  \left(\frac{M R_{AdS}^2}{u_\Lambda} \right)^2 
                     \frac{e^{2y}(1+e^y)^{p-3}}{(1+e^y)^{p+1}-1}. 
\end{equation}
For $p=4$ the WKB approximation to the $1^{-+}$-glueball spectrum is:

\begin{equation}   \label{eq:WKBV}
M^2_{\mathbf{1^{-+}}, h_{\theta \nu}} \approx  \frac{39.66}{\beta^2} k \left(k +
  \frac{9}{4} \right) + O(k^0).
\end{equation}
Table (\ref{TableV}) compares the WKB expression to the numerical results.

The scalar perturbation of the metric leads to a complicated  set of differential 
equation and here we will only quote the result of \cite{Constable:1999gb} for the WKB formula:

\begin{equation}  \label{eq:WKBS}
M^2_{\mathbf{0^{++}}, h_{\theta \theta}} \approx  \frac{39.66}{\beta^2} k \left(k +
  \frac{3}{2} \right) + O(k^0).
\end{equation}
Remarkably this expression reproduces the result (\ref{eq:WKBT}) for the $2^{++}$-glueballs.
This degeneracy, however, does not hold beyond the WKB approximation as one can see 
comparing the spectrum of the $2^{++}$ and $0^{-+}$-glueballs in Table (\ref{TableA})
and the results in Table (\ref{TableS}), where we present the numerical results for the 
$0^{++}$-glueballs.


\subsection{Comparison}

In Fig. \ref{TNCTL} we compare 
the  glueball spectrum for $YM_4$ in strong coupling
calculated above with the lattice spectrum \cite{Morningstar:1997ff} for pure $SU(3)$ $YM_4$.
In Fig. \ref{TC} we collect the results one obtains using the critical supergravity approach.
Few remarks are in order:

\FIGURE[tb]{
 \label{TNCTL}
\centerline{ \input{TableNonCritical.pstex_t} \input{TableLattice.pstex_t}}
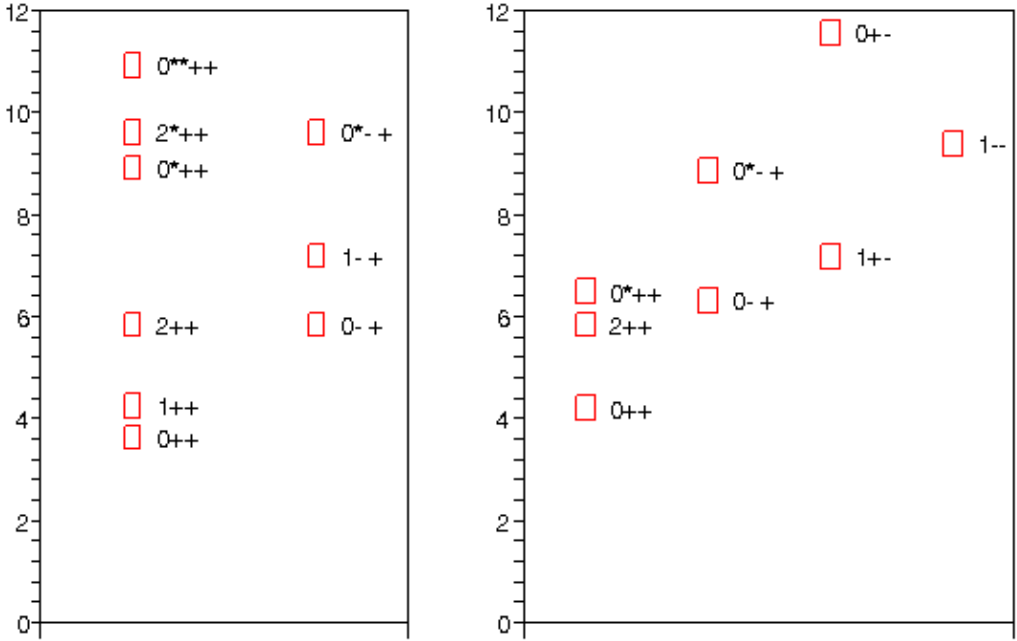
\caption{The AdS glueball spectrum for $YM_4$ computed in the
 framework of non-critical supergravity 
(left) and the corresponding lattice results (right). The AdS scale is adjusted to set the lowest 
$2^{++}$ state to the the lattice result 
in units of the hadronic scale $1/r_0 = 410 \textrm{MeV}$.  }}

\begin{itemize}
\item
  Although we did not succeed to reproduce accurate mass ratios for 
  all glueball states, we see that there is a remarkable similarity 
  between our strong coupling spectrum and the lattice results.
  For example, from the lattice computations one has $M(0^{-+})/M(0^{++}) \approx 1.504$, 
  while our calculation gave $M(0^{-+})/M(0^{++}) \approx 1.610$.
\item
   As in the critical supergravity calculations the lowest 
   $0^{++}$-glueball state comes from the scalar component of the metric, and not from
   the dilaton.
\item
  We saw above that the masses of the $2^{++}$ and the $0^{-+}$-glueball
  are identical even beyond the WKB approximation. Remarkably, in the lattice YM 
  spectrum the lowest masses of these states are quite close, more precisely 
  $M(2^{++})/M(0^{-+}) \approx 1.082$. In the critical supergravity computation
  the ratio is $M(2^{++})/M(0^{-+}) \approx 1.203$.

\FIGURE[tb]{
 \label{TC}
\centerline{\input{TableCritical.pstex_t}}
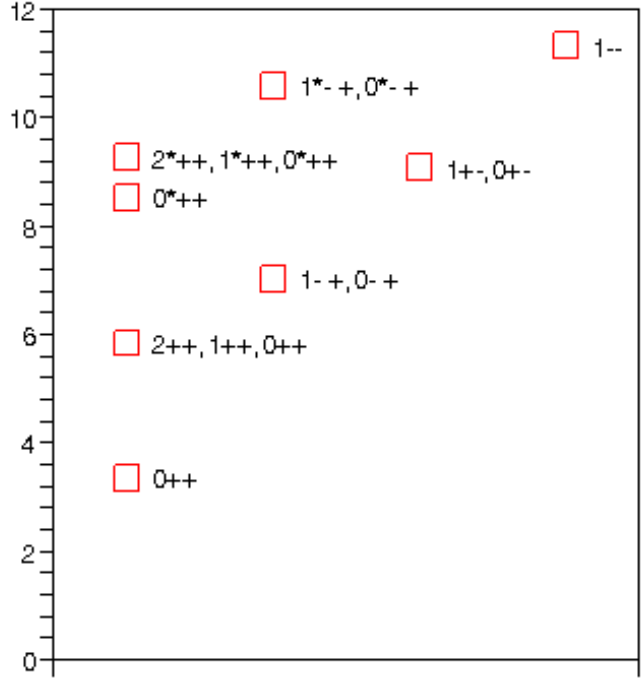
\caption{The AdS glueball spectrum for $YM_4$ computed in the
near extremal $AdS_7 \times S^4$ background with further reduction to the type IIA SUGRA solution. 
The AdS scale is adjusted to set the lowest 
$2^{++}$ state to the the lattice result in units of the 
hadronic scale $1/r_0 = 410 \textrm{MeV}$.  
}}

\item
In Witten's setup \cite{Witten:1998zw} one starts from the $AdS_7 \times S^4$ background and than
introduces two circles  $S^1 \times S^1$ with the  anti-periodic boundary conditions on one
of them and taking the radii to zero one reduces the world volume to four
dimensions.
In this limit $M$ theory reduces to type IIA string theory
on the non-conformal $D_4$ background with infinite temperature. 
Let us denote the coordinates of the 
$AdS_7$ black hole metric by $\theta$, $x_{i=1,2,3,4}$, $x_{11}$ and $u$.
Here $\theta$ and $x_{11}$ are the two compact directions and $u$ is the radial coordinates. 
The reduction to the type  IIA background 
corresponds to the compactification on $x_{11}$.
The graviton polarization tensor has $(5 \times 6)/2-1=14$ independent components.
It decomposes into $9$-dimensional tensor, $4$-dimensional vector
and $1$-dimensional scalar irreducible representation.
The graviton equation of motion, therefore, leads to three distinct wave equations
as we saw in the previous subsection.
This immediately implies that after dimensional reduction to $10d$
we obtain a \emph{degenerate} spectrum. In particular, the tensor wave
equation (\ref{eq:Tde})  for $p=5$ will lead to the $2^{++}$, $1^{++}$ and 
$0^{++}$ degenerate glueball spectrum in $d=4$. Similarly the vector equation
(\ref{eq:Vde}) for $p=5$ corresponds to the $1^{++}$ and 
$0^{++}$ glueballs.
There is no degeneracy between these states in our approach, since
we have only \emph{one} compact coordinate.
Moreover, there is a scalar field $h^\alpha_\alpha$ coming from the trace
of the metric on the $S^4$, which is related to the $0^{++}$-state in the gauge theory.
For this field the critical supergravity prediction is 
$m_{0^{++},h^\alpha_\alpha}/m_{2^{++}} \approx 2.28$.
In our model this field do not appear since there is no compact transversal space.

\item
For large $k$ the WKB expressions (\ref{eq:WKBphi}), (\ref{eq:WKBAtheta}),
(\ref{eq:WKBT}), (\ref{eq:WKBV}), (\ref{eq:WKBS}) for the glueball masses
reduce to 

\begin{equation}
M \approx \frac{C_p}{\beta} k,
\end{equation}
where $C_p$ is a constant depending on $p$. In the critical supergravity calculation ($p=5$)
one has $C_5 \approx 5.42$ 
and in our case ($p=4$) we get $C_4 \approx 6.30$.  It means that we obtain 
different asymptotic behaviors in our
non-critical description and Witten's model. Unfortunately, we 
cannot compare this result to the lattice YM, since it provides only masses of the lowest  
states.

\end{itemize}


\section { The Wilson loop} 
\label{TWl}

The stringy description of the Wilson loop \cite{Maldacena:1998im} \cite{Rey:1998ik} is in terms 
of a NG string whose end-points are  "nailed" at two points on the 
boundary of the AdS black hole space-time, namely at $u=\infty, x=\pm L/2$,
where $x$ denotes one of the 3d space directions. 
In \cite{Kinar:1998vq} the classical energy of the Wilson loop associated with a 
background with a general dependence on the radial direction was written down.
In particular it was proved that a sufficient condition for an area law behavior is 
that: 
\begin{equation}
 g_{uu}g_{00}(u_d)\rightarrow \infty 
 \qquad 
 \textrm{and}
 \qquad
 g_{00}(u_d)>0,
\end{equation} 
where $u_d$ is a particular point along the $u$ direction.
In this case the string tension is given by $T_s = g_{00}(u_d)$.

It is very easy to verify that our background metric (\ref{eq:bh}) obeys this condition
for $u_d=u_\Lambda$
and hence the potential of a quark anti-quark pair of the dual gauge theory 
is indeed linear in the separation distance $L$. 
In fact using the results of \cite{Kinar:1998vq}, we can write down the full
expression for the classical energy of the Wilson line. It takes for our case the following form:

\begin{equation}
E= \frac {1}{2\pi} \left ( \frac {u_\Lambda}{R_{AdS}}\right )^2 \cdot L - 2 \kappa
+ \mathcal{O} \left( ( \log L)^\gamma e^{-\alpha L} \right)
\end{equation}
where $\alpha= \sqrt{5} \frac{u_\Lambda}{R_{AdS}^2}$, $\gamma$ is a positive constant
and the constant $\kappa$ is given by:

\begin{equation}
\kappa =  \frac{1}{2 \pi}
 \int_{u_\Lambda}^{\infty} du 
  \left( \left( 1- \left( \frac{u_\Lambda}{u} \right)^{5} \right)^{-1/2}
        -1 \right) \approx 0.309 \frac{u_\Lambda}{2 \pi}.      
\end{equation}


\subsection{ The Luscher term}

The Luscher term is the sum of the contributions of  quantum fluctuations that adds to the 
classical quark anti-quark potential. 
In a superstring it incorporates both the bosonic as well as the fermionic fluctuations. 
In the present case of the non-critical string, 
we lack a  formulation of the fermionic part of the action and in fact 
it is plausible that even prior to invoking the anti-periodic
boundary condition along the thermal circle, 
the model is not space-time supersymmetric. 
Hence, we discuss here only the bosonic quantum fluctuations.  
However, it might be that in spite of this ignorance 
we can predict the form of the full contribution of the 
quantum fluctuations. 
Recall  that in the critical case, due to the coupling to the RR fields, all the fermionic 
modes are massive so that in that case the contributions of the massless modes are coming only
from the bosonic sector. 
Since in our non-critical model we also have RR fields it is quite
plausible that a similar mass generation for the fermions will take place.  

The fluctuations along the bosonic directions fall into two classes:
massless modes and massive modes.
The latter contribute to the quark anti-quark potential a Yukawa like
term $\sim e^{-mL}$, 
where $m$ is the mass of the mode. Since in computing Wilson loops we
take $L\rightarrow \infty$, the contribution of  massive
modes is negligible. The contribution of a massless mode has the form 
$-\frac{1}{24} \frac{\pi}{L}$. Thus what is left to be determined is the number of massless modes.
It turns out that this issue, which is very crucial when comparing to
the lattice result, is a subtle one and 
there are contradicting claims about it in the literature. 
In \cite{Greensite:1999jw} \cite{Bigazzi:2004ze} 
in the context of the critical theory it was found that there are 
two massive bosonic modes. Applying  this result  to our case it will
imply that there 
are altogether
$6-2-2=2$ massless modes and hence  $\Delta E_B \sim -\frac{2}{24} \frac{\pi}{L} $.
It is very tempting to adopt this result since it 
agrees with the current value found in lattice calculations.
However, following \cite{Kinar:1999xu} we claim that in fact the
number of massive modes in one and hence we 
 end up with a Luscher term:

\begin{equation}
\Delta E_B =-\frac{3}{24} \frac{\pi}{L} 
\end{equation} 
whereas in the critical case the result was $\Delta E_B =-\frac{7}{24} \frac{\pi}{L} $.
Obviously even without using the results of  \cite{Greensite:1999jw}
\cite{Bigazzi:2004ze}, 
the outcome  of  the non-critical string model
is closer to the value measured in the lattice calculations than the 
result of the critical string model. 

Since for  the comparison with lattice simulations, 
the difference between the result of  \cite{Greensite:1999jw}
\cite{Bigazzi:2004ze} 
and our claim is important, let us briefly comment
about the source of the discrepancy following
the derivation of \cite{Kinar:1999xu}. The idea there is to fix the
gauge of the NG action by $\tau=t, \sigma=u_{cl}$
so that the fluctuation in the $(x,u)$ plane is along the normal to the classical configuration. 
It was shown that choosing this gauge avoids potential problems 
of the $\sigma=u$ and $\sigma=x$ gauges.
It was further shown that one of the modes that was found in
\cite{Greensite:1999jw} to be massive must be massless since
it corresponds to a Goldstone boson associated with a breaking of a
rotation invariance \cite{Kinar:1999xu}.


\subsection{ The correlator of two Wilson loops}

The  correlator of two Wilson loops in the context of the $AdS_5\times S^5$ model 
was discussed in \cite{Gross:1998gk} \cite{Zarembo:1999bu}. 
It was shown  that when the separation of the
two loops in AdS$_{5}$ is of the order of their size, there is a solution of the
string equation of motion that describes a connected surface ending on
the two loops. As the separation increases the tube
shrinks, and becomes unstable. 
At large distances the correlation is due to the exchange of
supergravity modes in the bulk between the world-sheets of the loops. 
The  long distance correlator was calculated in  \cite{Berenstein:1998ij}.
An analysis of the long distance correlator for the near extremal
$AdS_5\times S^5$ model 
in the limit of large temperature that corresponds to the 
low energy effective action of the pure three dimensional 
YM theory was performed in \cite{Sonnenschein:1999re}.
It was shown that the correlator was dominated by an exchange of a KK
``scalarball''  mode which is lighter than any of the glueball modes.

Let us now address the correlator for our  non-critical near extremal $AdS_6$ model.
It is plausible that the transition between a connecting world-sheet 
into an exchange of a SUGRA modes will take place here also. 
For separation distances much greater than the distance 
between the endpoints of each string, the correlator is given by:

\begin{equation} \label{cor} 
\log \left[ \frac{ \left< W(0)W(R) \right>}{ \left< W(0) \right> \left<W(R) \right>} \right] 
    = \sum_{i,k} \int {\cal A}_{1} \int {\cal A}_{2}
 \ f_{1}^{i,k} \  f_{2}^{i,k}  \ G^{i,k},
\end{equation}
where $R$ is the distance between the two loops, ${\cal A}_{i}$ are
the areas of the two loops, 
$k$ is the mode number and 
 $f_{1}^{i,k}, \ f_{2}^{i,k}$ are the couplings of the field $i$
to the world-sheet and $G^{i,k}$ is the propagator.
There are two clear differences between this expression 
and the corresponding one in the critical case:
(i) The sum of the modes here does not include any of the KK modes 
associated with the $S^5$ and hence only glueballs 
(and modes associated with the thermal $S^1$). 
(ii) For each  of the modes like the dilaton, the graviton etc. 
we do not sum over all the spherical harmonics associated with the $S^5$.
The dilaton coupling follows
simply from the relation between the Einstein and string frame
metrics, $g_{s}=g_{E}e^{\phi}$. Therefore the one dilaton coupling
is $1$. The coupling to the other modes are in general $u$ dependent 
\cite{Sonnenschein:1999re}.

There are at least three types of correlators that one can study: 
(i)  Correlators of circular loops located on  planes in the space
spanned by the $x^i$ coordinates. 
(ii) Correlators of infinite strips 
along $(t, x^i)$ planes (iii) Wilson loops along $(\theta, x^i)$. 
The first two types where discussed first
for the  $AdS_5\times S^5$ string in \cite{Berenstein:1998ij} and in
\cite{Sonnenschein:1999re} the three types were discussed in 
the near extremal case which is  dual to the low energy regime of
$3d$ pure YM theory (contaminated with KK modes).
In a similar way the correlator of type (ii) in our non-critical model
should correspond to the potential between two external mesons of $4d$
pure YM theory.  
The result we find for this potential is:

\begin{equation}  \label{pot} 
V_{mm} \sim \frac{1}{T} \log \left[
\frac{ \left<W(0)W(R) \right>}{ \left<W(0) \right> \left<W(R)\right>}\right] \sim 
         \frac {1}{N^{2}} K_{0}(M_{\varphi_{00}}R) ,
\end{equation}
where $T$ is the temporal length of each strip, 
$K_{0}$ is a modified Bessel function and $\varphi_{00}$ is the lightest scalar glueball mode. 
Based on the analysis of Section \ref{Tgs}, 
this mode will be the $0^{++}$ that associates with the graviton.
Note that the amplitude scales like $\frac{1}{N^2}$. 
This is also the behavior in the critical case. In the latter case the
amplitude is also proportional to $g_{YM}^2 N$ 
which relates to the radius of the $AdS_5$ and hence does not show up in our formulation.


\section { The 't Hooft loop} 
\label{TtHl}

It is well known that in confining gauge theories a monopole 
anti-monopole are screened from each other. The corresponding potential is determined via
the 't Hooft loop in a similar way that the potential between a quark and anti-quark is
determined  from the Wilson loop.
The stringy description of a monopole is a $D2$ brane that ends on a D$4$ brane 
so that the 't Hooft loop is described by a  $D2$ brane that is attached to the 
boundary of the background in two points.
In the context of critical string theories the 't Hooft loop was determined in 
\cite{Brandhuber:1998er} for the case of the near extremal D$4$ brane background, 
and it was shown that indeed in that case the system 
of a monopole and anti-monopole was  energetically  favorable 
in comparison with the bound state system, which implies that 
the monopole and anti-monopole are indeed screened from each other.
 
Let us perform a similar calculation for our setup. 
We take the world-volume of the  $D2 $ brane to be 
along the directions $t$, $x$ and $\theta$, namely it wraps the $S^1$.
We will assume that  the radial coordinate $u$ along the $D2$ brane depends only on 
$x$.
The action of the $D2$ brane takes the form:

\begin{equation}
S = \frac{1}{(2\pi\alpha')^{3/2}} 
    \int d\tau d \sigma_1 d \sigma_2 e^{-\phi_0} \sqrt{ \textrm{det} h_{\textrm{ind}} }  
  =   \beta  \int_{-L/2}^{L/2} dx \frac {u}{ R_{AdS}} 
          \sqrt { {u^\prime}^2 + \left( \frac{u}{ R_{AdS}} \right)^4 f(u) } ,   
\end{equation}
where $f(u)=1-(u_\Lambda/u)^5$ is the thermal factor
and $L$ is the separation distance between the pair
since the action does not depend on $x$ explicitly 
the solution of the equation of motion satisfies:

\begin{equation}
\left( \frac{u}{R_{AdS}} \right)^5 f(u)
     \left( {u^\prime}^2 + \left( \frac{u}{ R_{AdS}} \right)^4 f(u) \right)^{-1/2} =
 \textrm{const}.      
\end{equation}
Defining $u_{\textrm{min}}$ to be the minimal value $u(x)$ we arrive at the following 
expression for the separation distance $L$:

\begin{equation}
L = \int dx = 2 \int _{u_\Lambda} ^\infty  \frac{du}{u^\prime} = 
2 \frac{R^2_{Ads}}{u_{\textrm{min}}} \epsilon^{1/2} 
 \int_1^\infty dy
  \frac{y}{ \sqrt{(y^5-1+\epsilon) (y^5 -\epsilon y -1+\epsilon)} },
\end{equation}
where $\epsilon \equiv f(u_{\textrm{min}}) $.
For $u_{\textrm{min}} \to u_\Lambda$ we have $\epsilon \to 0 $ and $L \to \infty$.
To compute the binding energy we have to subtract the 
masses of the free monopole and anti-monopole, namely the energy
of two parallel $D2$ branes stretched from the boundary to the horizon.
The final result is:

\begin{equation}  \label{eq:tHooftEnergy}
\Delta E \sim  \beta \frac{u_{\textrm{min}}^2}{u_\Lambda} \left[
 \int_1^\infty dy \, y^2
   \left( \sqrt{ \frac{y^5-1+\epsilon }{ y^5 -\epsilon y -1+\epsilon } } -1 \right)
 -\frac{1}{3} \left( 1 - \left( \frac{u_\Lambda}{u_{\textrm{min}} } \right)^3 \right)
    \right]  . 
\end{equation}
For  $L \beta^{-1} \gg 1 $ ($\epsilon \approx 0$) the energy is positive 
which means that the zero-energy configuration of two parallel $D2$ branes ending on the horizon
is energetically favorable.  We conclude therefore that in the "YM region" there is
no force between the monopole and the anti-monopole
and there is a screening of the magnetic charge.

Finally we should note that though our final conclusion matches the results of 
\cite{Brandhuber:1998er}, the 
explicit expression for the energy (\ref{eq:tHooftEnergy}) differs from 
the expression derived in \cite{Brandhuber:1998er}.


\section{Closed spinning strings in the $AdS_{p+1}$ black hole background}

\label{Css}

The glueball spectrum that was extracted from the supergravity in section 2 is
obviously limited to states of spin not higher than two. To study the spectrum
of glueballs of higher spin one has to consider stringy configurations rather than  
supergravity  modes.  The string configurations which are dual to glueballs of 
higher spin are spinning folded closed strings \cite{Gubser:2002tv}.
In particular we would like to investigate the possibility that the high spin glueballs
furnish  a close string Regge trajectory.
Our task is to check whether  the non-critical strings associated with the  $AdS_6$ 
black hole admit 
classical spinning configurations, compute the relation between the 
angular momentum and energy of such
configurations and incorporate quantum fluctuations.
This type of analysis was done previously in the context of  AdS black hole \cite{Armoni:2002fr},
in the framework of the KS and MN models in
\cite{PandoZayas:2003yb} and recently for the critical near extremal
D4 brane in \cite{Bigazzi:2004ze}.

Let us analyze now what are the condition of having 
classical spinning string solutions of the equations of motion.
Since the background depends  only on the radial coordinate $u$, the  
equation of motion with respect to this coordinate plays a special
role. 
Suppose now that we perform a coordinate transformation  $u
\rightarrow \rho(u)$, 
then the equation of motion with respect to $\rho$ in the Polyakov formulation reads:
\begin{equation}\label{eomrho}
-2\pa_\alpha ( g_{\rho\rho}\pa^\a \rho)  + \frac{d g_{\rho\rho}}{d\rho}  \pa_\a \rho \pa^\a \rho
-\frac{d g_{00}}{d\rho}  \pa_\a t \pa^\a t 
+\frac{d g_{ii}}{d\rho}  \pa_\a x^i \pa^\a x^i +
\frac{d g_{\theta\theta}}{d\rho}  \pa_\a \theta \pa^\a \theta 
= 0
\end{equation}  
It is trivial to check that a  spinning string of the form:

\begin{eqnarray}    \label{eq:foldedstring}
 &X^0 = e \tau \qquad
X^1 = e \cos \tau \sin \sigma \qquad
X^2 = e \sin \tau \sin \sigma  &   \nonumber \\
&\theta = \textrm{const}  \qquad   \qquad
\rho = \textrm{const} ,& 
\end{eqnarray}
solves the other equations of motion and that for such a configuration  
the first two terms of (\ref{eomrho}) and the last   vanish. 
Since the configuration has non trivial $X^0$ and $X^i$, (\ref{eomrho})
 is obeyed if at a certain value of $\rho=\rho_0$
$\frac{d g_{00}}{d\rho}|_{\rho=\rho_0}=\frac{d g_{ii}}{d\rho}|_{\rho=\rho_0}=0$.
It is natural to make  a coordinate transformation such that around
the horizon:

\begin{equation}
g_{uu}du^2 = d\rho^2 \rightarrow \qquad \frac{d \rho}{d u} = \frac{R_{AdS}}{u} f^{1/2}(u).
\end{equation}
In the new coordinates  $u_\Lambda$, namely the horizon, is mapped into
 $\rho=0$  as can be seen from

\begin{equation}
\rho \approx \frac{2 R_{AdS}}{\left(5 u_\Lambda \right)^{1/2}} (u-u_\Lambda)^{1/2}
 = \frac{\beta}{2 \pi} (u-u_\Lambda)^{1/2},
\end{equation}
and for the metric:

\begin{equation}
-g_{00}  = g_{ii} \approx \left( \frac{u_\Lambda}{R_{AdS}} \right)^2 + 
    \frac{5 u_\Lambda^2}{2 R_{AdS}^4} \cdot \rho^2 + \ldots.
\end{equation}
It implies that:
\begin{equation}    \label{eq:condition}
 g_{00} \vert_{\rho=0,u=u_\Lambda} \neq 0
 \qquad
 \textrm{and}
 \quad
 \pa_\rho g_{00} \vert_{\rho=0,u=u_\Lambda} = 0.
\end{equation}
which means that indeed for a spinning configuration at $\rho=0$ the equation of motion
(\ref{eomrho}) is obeyed.
In fact this is precisely one of the two possible sufficient 
conditions to have an area law Wilson loop \cite{Kinar:1998vq}.
Moreover, it implies that the spinning string stretches along the 
horizon which is often referred to as the 
``wall'' or the `` end of the world floor'' exactly as the static configuration of the string that
corresponds to the Wilson loop does. 
This relation between the condition for confinement and for spinning string configurations 
and the fact that the string, like the Wilson loop string, is along the wall,  
was observed also
in \cite{PandoZayas:2003yb} for folded close string 
in the KS and MN models and for open strings in the near extremal D4 brane model
in \cite{PandoZayas:2003yb}.  
It is very plausible that this relation is universal and 
applies to any SUGRA dual of a confining gauge theory.

It is also easy to check that the classical configuration (\ref{eq:foldedstring}) 
at the wall $\rho=0$ obeys 
on top of the equations of motion also the Virasoro constraint. 
For this solution the classical energy of the string configuration
is:

\begin{equation}
E =  \frac{e}{2 \pi \alpha^\prime} \int g_{00} \vert_{\rho=0} d \sigma = 
  2 \pi T_{\textrm{s}} e g_{00} \vert_{\rho=0}.
\end{equation}
Similarly the classical angular momentum is:

\begin{equation}
J =  \pi T_{\textrm{s}} e^2 g_{ii} \vert_{\rho=0}.
\end{equation}
Recalling that $g_{00} (0) = g_{ii} (0) \neq 0$ we end up with the following
result:

\begin{equation}
J = \half \alpha^\prime_{\textrm{eff}} E^2 = \half \alpha^\prime_{\textrm{eff}} t
\qquad
\textrm{where}
\qquad
 \alpha^\prime_{\textrm{eff}} = \frac{\alpha^\prime}{g_{00}(0)} = 
  \frac{\alpha^\prime}{\alpha^\prime\left( \frac{u_\Lambda}{R_{AdS}} \right)^2}
      = \frac{15/2}{u_\Lambda^2} .
\end{equation}


\subsection{ Quantum corrections- deviations from linearity}

It is well known that the basic linear relation between the 
angular momentum and $E^2\equiv t$ receives corrections.
The ``famous'' correction is the intercept $\alpha_0$  
which is a constant $t$ independent so that once included the 
trajectory reads $\half \alpha^\prime_{\textrm{eff}} t + \alpha_0$. 
In the string derivation of the Regge trajectory the intercept 
is a result of the quantum correction and hence it is 
intimately related to the Luscher term.   
For instance in the bosonic string in $d$ dimensions the intercept is $(d-2)\frac{\pi}{24}$.
This result is achieved by adding quadratic fluctuations to the classical configurations and 
``measuring'' the impact of these fluctuations on $J$ and $E$.
It turns out that in the canonical quantization of the 
Polyakov formulation, using the Virasoro constraint, 
one finds that \cite{Tseytlin:2003ac}: 

\begin{equation}
e(E-\bar E) = J- \bar J + \int d\sigma {\cal H}(\delta x^i), 
\end{equation}
where $\bar E$ and $\bar J$ are the classical values of the energy and angular momentum and 
${\cal H}(\delta x^i) $ is the world-sheet Hamiltonian expressed in terms of the fluctuating fields.
Using this procedure, as well as a path integral calculation,
the contributions of the quantum fluctuations  were computed  
in the KS and MN models \cite{PandoZayas:2003yb}
and in the near near extremal $D4$ model \cite{Bigazzi:2004ze}.
From the  result of the latter one can be easily extract the form of the result also for our
non-critical model. There are two apparent differences, though: (i)  The number of physical 
bosonic directions which is here 4 rather 8, (ii) The fact that we do
not have a formulation for the fermionic fluctuations in our model. 
Hence, we can only determine the contribution of
the bosonic fluctuations to the non linearities of the trajectory. 
The bosonic modes include $3$ massless modes and one massive mode.
The latter has a $\sigma$-dependent mass.
Its contribution was calculated in \cite{Bigazzi:2004ze} \cite{PandoZayas:2003yb}.
Collecting all the contributions we get:

\begin{equation}
J = \half \alpha^\prime_{\textrm{eff}}(E-z_0)^2 - \frac{3}{24} \pi + \Delta_{\textrm{f}},
\end{equation}
where $z_0$ is proportional to $u_\Lambda$ and $\Delta_{\textrm{f}}$
will be the contribution of the massless and massive fermionic modes.
We thus see that there is a non trivial bosonic intercept 
$\alpha_0= \half \alpha^\prime_{\textrm{eff}}z_0^2 - \frac{3}{24} \pi$, 
but in addition there is also a term linear in $E$.
A difference between the result in the critical model 
\cite{Bigazzi:2004ze} and the non-critical one is in the proportionally
factor between $z_0$ and $u_\Lambda$.

\acknowledgments

We would like to thank Ido Adam, Yaron Oz and Ofer Aharony for
fruitful discussions.  
J.S. would like to thank Leo Pando Zayas and Diana Vaman.
He also wants to thank the Kavli Institute for Theoretical Physics
at UC, Santa Barbara and the Department of Physics
of the University of Texas at Austin, where part of this work was done.
This work was supported in part by the German-Israeli Foundation for
Scientific Research and by the Israel Science Foundation.

\appendix

\bibliography{xx}

\end{document}

%% file: Vdilaton20.pstex_t
\begin{picture}(0,0)%
\includegraphics{Vdilaton20.pstex}%
\end{picture}%
\setlength{\unitlength}{4144sp}%
\begingroup\makeatletter\ifx\SetFigFont\undefined%
\gdef\SetFigFont#1#2#3#4#5{%
  \reset@font\fontsize{#1}{#2pt}%
  \fontfamily{#3}\fontseries{#4}\fontshape{#5}%
  \selectfont}%
\fi\endgroup%
\begin{picture}(5000,5000)(1576,-5324)
\put(6326,-3549){\makebox(0,0)[b]{\smash{{\SetFigFont{14}{16.8}{\rmdefault}{\mddefault}{\updefault}$y$}}}}
\put(3826,-874){\makebox(0,0)[rb]{\smash{{\SetFigFont{14}{16.8}{\rmdefault}{\mddefault}{\updefault}$V(y)$}}}}
\put(4526,-3774){\makebox(0,0)[b]{\smash{{\SetFigFont{14}{16.8}{\rmdefault}{\mddefault}{\updefault}$y_+$}}}}
\end{picture}%

%% file: Vrr5.pstex_t
\begin{picture}(0,0)%
\includegraphics{Vrr5.pstex}%
\end{picture}%
\setlength{\unitlength}{4144sp}%
\begingroup\makeatletter\ifx\SetFigFont\undefined%
\gdef\SetFigFont#1#2#3#4#5{%
  \reset@font\fontsize{#1}{#2pt}%
  \fontfamily{#3}\fontseries{#4}\fontshape{#5}%
  \selectfont}%
\fi\endgroup%
\begin{picture}(5000,5000)(1576,-5324)
\put(3826,-874){\makebox(0,0)[rb]{\smash{{\SetFigFont{14}{16.8}{\rmdefault}{\mddefault}{\updefault}$V(y)$}}}}
\put(3339,-3374){\makebox(0,0)[lb]{\smash{{\SetFigFont{14}{16.8}{\rmdefault}{\mddefault}{\updefault}$y_-$}}}}
\put(6339,-3699){\makebox(0,0)[b]{\smash{{\SetFigFont{14}{16.8}{\rmdefault}{\mddefault}{\updefault}$y$}}}}
\put(4289,-3874){\makebox(0,0)[lb]{\smash{{\SetFigFont{14}{16.8}{\rmdefault}{\mddefault}{\updefault}$y_+$}}}}
\end{picture}%

%% file: TableNonCritical.pstex_t
\begin{picture}(0,0)%
\includegraphics{TableNonCritical.pstex}%
\end{picture}%
\setlength{\unitlength}{4144sp}%
\begingroup\makeatletter\ifx\SetFigFont\undefined%
\gdef\SetFigFont#1#2#3#4#5{%
  \reset@font\fontsize{#1}{#2pt}%
  \fontfamily{#3}\fontseries{#4}\fontshape{#5}%
  \selectfont}%
\fi\endgroup%
\begin{picture}(2743,4572)(789,-4446)
\end{picture}%

%% file: TableLattice.pstex_t
\begin{picture}(0,0)%
\includegraphics{TableLattice.pstex}%
\end{picture}%
\setlength{\unitlength}{4144sp}%
\begingroup\makeatletter\ifx\SetFigFont\undefined%
\gdef\SetFigFont#1#2#3#4#5{%
  \reset@font\fontsize{#1}{#2pt}%
  \fontfamily{#3}\fontseries{#4}\fontshape{#5}%
  \selectfont}%
\fi\endgroup%
\begin{picture}(3657,4572)(601,-4321)
\end{picture}%

%% file: TableCritical.pstex_t
\begin{picture}(0,0)%
\includegraphics{TableCritical.pstex}%
\end{picture}%
\setlength{\unitlength}{4144sp}%
\begingroup\makeatletter\ifx\SetFigFont\undefined%
\gdef\SetFigFont#1#2#3#4#5{%
  \reset@font\fontsize{#1}{#2pt}%
  \fontfamily{#3}\fontseries{#4}\fontshape{#5}%
  \selectfont}%
\fi\endgroup%
\begin{picture}(4371,4857)(789,-4731)
\end{picture}%